\documentclass[referee]{raa}            

\usepackage{graphicx,times}             
\usepackage{natbib}
\usepackage{amssymb,amsmath}
\usepackage[T1]{fontenc}
\bibpunct{(}{)}{;}{a}{}{,}

\usepackage[a4paper=true,dvipdfm=true,pagebackref=true]{hyperref}
\hypersetup{colorlinks = true, linkcolor = green, anchorcolor = red, citecolor = blue, filecolor = red, pagecolor = red, urlcolor = red}

\begin{document}

   \title{Age and chemical composition of the globular cluster NGC~6652
}
   \volnopage{Vol.20 (2020) No.128, 000--000}      
   \setcounter{page}{128}          

   \author{Margarita Eugene Sharina
      \inst{1}
   \and Vladislav Vladimirovich Shimansky
      \inst{2}
   }

   \institute{Special Astrophysical Observatory, Russian Academy of Sciences,
Nizhnii Arkhyz, 369167, Russia; {\it sme@sao.ru}\\
        \and
            Kazan Federal University, 18 Kremlyovskaya street, Kazan, 420008, Russia \\
\vs\no
   {\small Received~~2020 January 23; accepted~~2020~~March 10}}

\abstract{We present the results of determination of the age, helium mass fraction (Y), metallicity ([Fe/H]),
and abundances of the elements $ \rm C$, $ \rm N$, $ \rm O$, $ \rm Na$, $ \rm Mg$, $ \rm Ca$, $ \rm Ti$, $ \rm C$ and $ \rm Mn$ for the Galactic globular cluster 
NGC~6652. We use its medium-resolution integrated-light spectrum from the library of Schiavon and our population
synthesis method to fulfill this task. We select the evolutionary isochrone and stellar mass function for our analysis,
which provide the best approximation to the shapes and intensities of the observed Balmer line profiles.
The determined elemental abundances, age and metallicity are characteristic of stellar populations in the Galactic Bulge.
}

   \authorrunning{M. E. Sharina \& V. V. Shimansky }            
   \titlerunning{Age and chemical composition of NGC~6652}  

   \maketitle

%
%
\section{Introduction}           
\label{sect:intro}
NGC~6652 is an intermediate metallicity 
($\rm [Fe/H]=-0.96$ dex and $\rm [Fe/H]=-0.85$ dex in the \cite{ZinnWest84} and \cite{CarrettaGratton97}
scales, respectively)\footnote{
The iron content in solar units is $\rm[Fe/H] = log(N_{Fe}/N_H)-log(N_{Fe}/N_H)_{\sun}$, where $ \rm N_{Fe}/N_H$
is the ratio of the abundances of iron and hydrogen in terms of numbers of atoms, or in
terms of mass, which is related to the mass fraction of elements heavier than helium (Z). 
The solar mass fractions of H - $X$, He - $ \rm Y$, and metals $ \rm Z$ are given in \cite{Asplund}. Obviously, X + Y + Z = 1.},
old Galactic globular cluster (GC, 11.7~Gyr, \citealt{Chaboyer00}). It is at the heliocentric distance of 10~kpc and is mildly obscured by Galactic dust
 ($\rm E(B-V)\sim 0.1$~mag) (\citealt{Bica16} and references therein).
It is thought to be associated with the bulge (\citealt{Bica16}) or with the inner halo (\citealt{Chaboyer00}). 

 We did not find elemental abundances obtained in high-resolution spectroscopic studies for stars in NGC~6652 in the literature.
\cite{Conroy} estimated the relative age (1.01$\pm0.01$), $\rm [Fe/H]=-0.93\pm0.02$~dex and abundances of C, N, Mg, Si, Ca and Ti for
NGC~6652 using the integrated-light (IL) spectrum from \cite{Schiavon05} and their updated stellar population models. 
 \cite{Schiavon05} found $\rm [Fe/H]=-1.1$~dex using this spectrum (their Table~1).

 The elongated orbit with the apocenter radius of 8.1~kpc and eccentricity of 0.58 (\citealt{Balbinot18}), proximity to the Galactic
center and rather low metallicity make NGC~6652 an interesting object of study in the context of Galactic evolution.

In this work we make use the integrated spectrum of NGC~6652 from \cite{Schiavon05} 
and our population synthesis method (\citealt{SS19}, \citealt{SSK18}, \citealt{SSK17} and references therein) 
to derive the helium mass fraction (Y), age and abundances of several chemical elements. 

\section{Observational data}
\label{sect:data} 
The spectrum of NGC~6652 was obtained by \cite{Schiavon05} with the Ritchey-Chr{\'e}tien spectrograph mounted on the 4 m Blanco telescope at the 
Cerro Tololo Inter-American Observatory (CTIO) with the $1.5\arcmin \times5.5\arcmin$ long slit oriented 
in the East-West direction. The  wavelength range is 3360-6430~\AA\ and  the spectral resolution is 3.1~\AA\
full width at half maximum ($\rm FWHM$)
in the central $\sim1000$\AA, deteriorating to 3.6~\AA\ $\rm FWHM$ at the blue and red ends of the spectrum.
According to the adopted strategy of the observations, the spectrum was obtained by drifting the 
spectrograph slit across the core diameter of the cluster ($\pm 9 \arcsec$ from the center of NGC~6652).
The background spectrum was obtained at the angular offset from the cluster center of 12\arcmin\ to the West.
The spectrograph slit was trailed within the angular distance of 5\arcmin\ for the sky background exposure.

The quality of the IL spectrum of NGC~6652 from \cite{Schiavon05} is very good (please,
see the original paper for the complete description of the observations, instrumental setup and spectra reduction).
Although the authors do not report the resulting signal-to-noise ratio ($ \rm S/N$), they compare spectra of 
 two moderately metal-poor clusters with similar metallicity ($ \rm [Fe/H]\sim-1.1$~dex), NGC~6723 and NGC~6652, 
 and demonstrate in their Fig.~3, that the number of stars in the blue part of the horizontal branch (HB) may influence 
 the intensities of absorption lines in the spectra. Balmer lines are weaker and metal lines are stronger in the spectrum of
 NGC~6652 with the redder HB. These definite conclusions would not be possible from the visual comparison of the spectra with poor $ \rm S/N$.
 As will be shown in the next section, our analysis confirms that the $ \rm S/N$ in the spectrum is high.
 The probability of a noticeable contribution of Galactic field stars in the spectrum is low, despite the fact that NGC~6652
 is located in a crowded stellar region.

\section{Data analysis and results}
\label{sect:result}

The spectrum of NGC~6652 was compared with the synthetic IL spectrum computed according to 
effective temperatures $ \rm T_{eff}$, surface gravities $ \rm log(g)$ and metallicities $ \rm [Fe/H]$ 
set by theoretical evolutionary isochrones for stars in the cluster.
In this work, we use the scaled-solar isochrones by \citealt{Bertelli+08} (hereafter: B08)
and by the {\textsc{Teramo}} group (\citealt{Pietrinferni+13} and references therein, hereafter: 
{\textsc{Teramo}} isochrones). 

Synthetic spectral calculation
is based on the plane-parallel, hydrostatic stellar atmosphere models by \cite{CastelliKurucz}\footnote{ 
The lists of atomic and molecular lines are adopted from the R.L. Kurucz web
site (http://kurucz.harvard.edu/linelists.html).}.
 The calculated synthetic spectra of individual stars were summed according to the selected mass function (MF). 
In this work, we apply the MFs by \cite{Chabrier05} (equation (2) in that paper, hereafter: Ch05) 
and \cite{Salpeter55} (please see the paper
Ch05 for a discussion about the relation between the initial and present-day MFs). 
\begin{figure}[h]
   \centering
   \includegraphics[bb=53 30 570 800,width=6.7cm, angle=-90,clip]{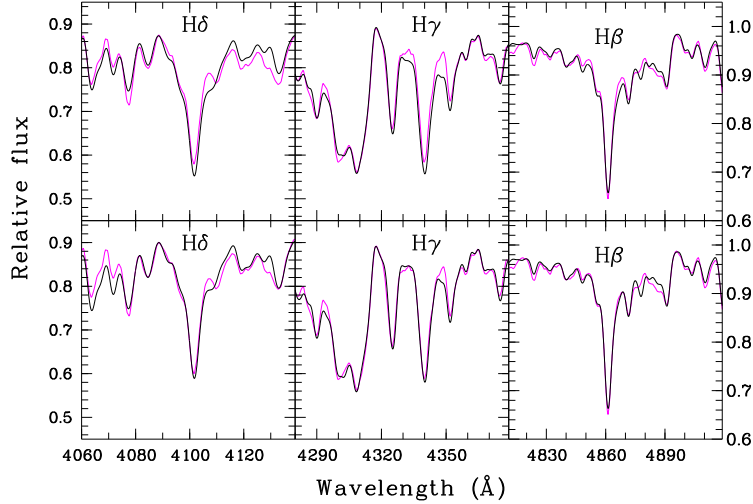}
   \caption{{\small Comparison of three Balmer lines in the continuum normalized spectrum of NGC~6652 (\citealt{Schiavon05}) (magenta) 
   and the synthetic IL spectra (black) computed using  the MF by Ch05, 
   the same elemental abundances but different isochrones:
   Z=0.002, Y=0.26, log(age)=10.15 B08 (bottom) and Z=0.002, Y=0.248, age=14 Gyr 
   ({\textsc{Teramo}}) (top).}}
   \label{Fig:result}
   \end{figure}
Let us note that the preferred isochrone and the MF can be selected purely on the basis of
comparing the strengths and shapes of Balmer lines in the observed and synthetic spectra.

The best correspondence between observational and theoretical data is illustrated in the bottom panels 
of Fig.~\ref{Fig:result}. It can be seen that the synthetic spectrum computed with
the isochrone $ \rm Z=0.002$, $ \rm Y=0.26$, $ \rm log(age)=10.15$ (B08) and the Ch05 MF better
describes the observed Balmer lines than the spectrum computed with the {\textsc{Teramo}} isochrone
$ \rm Z=0.002$, $ \rm Y=0.248$, $\rm age=14$~Gyr. 

\begin{figure}[h]
   \centering
       \includegraphics[bb=30 15 570 500,width=65mm, angle=-90,clip]{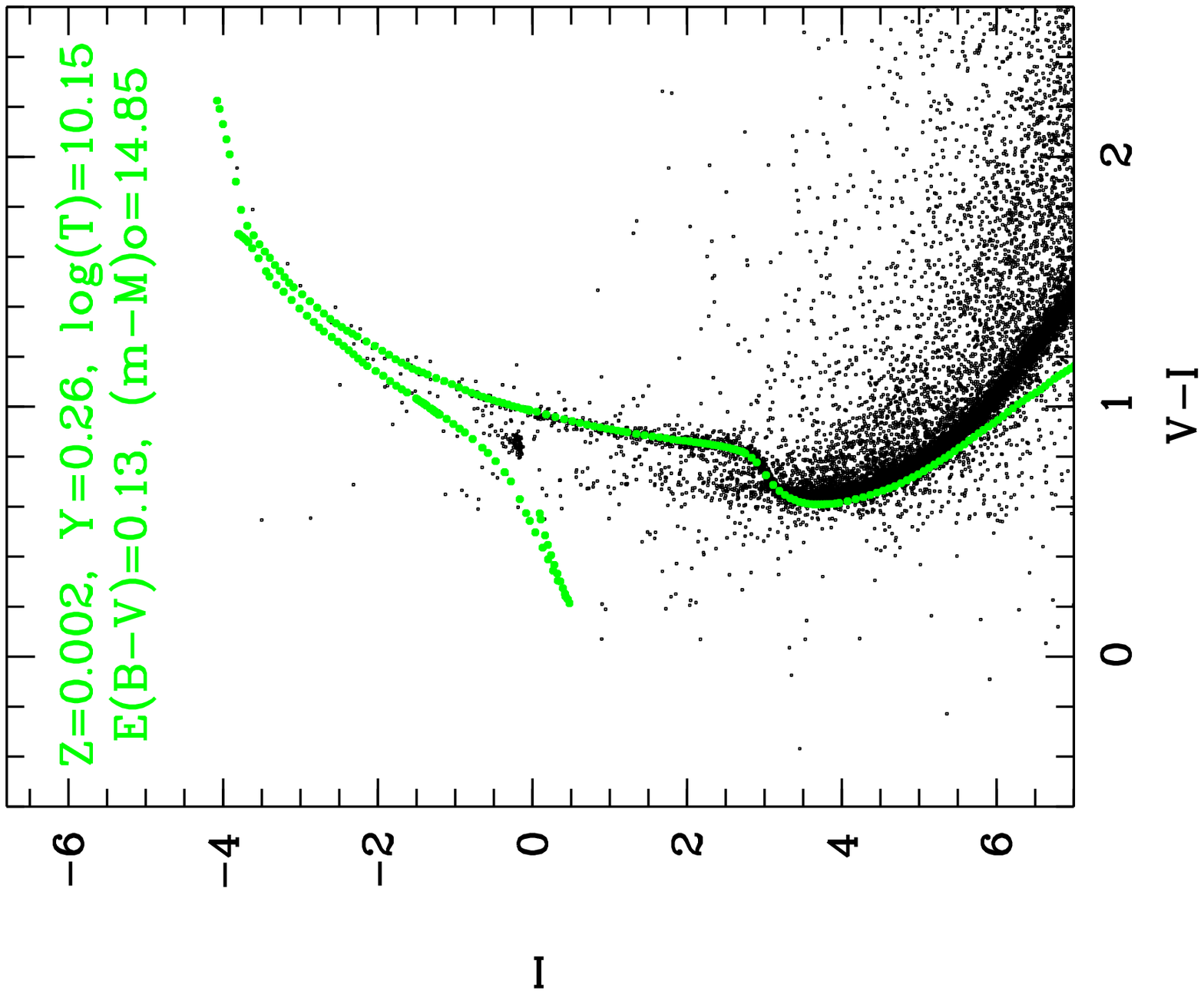}
       \includegraphics[bb=30 35 570 500,width=65mm, angle=-90,clip]{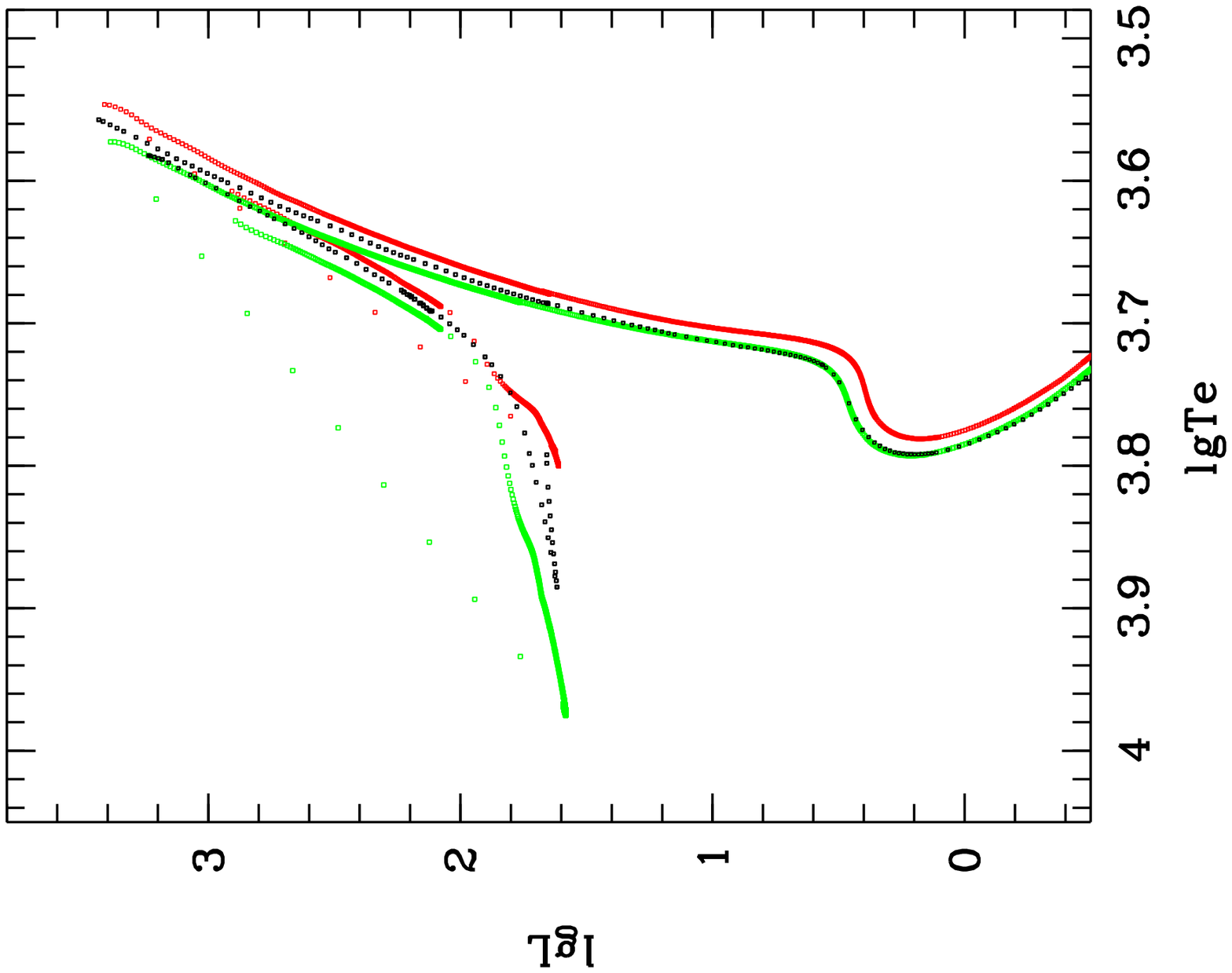}
       \caption{{\small Left: comparison between the CMD of NGC6652 (\citealt{Sarajedini07}) 
       and the theoretical evolutionary isochrone Z=0.002, Y=0.26, log(age)=10.15 (B08).
       Right: comparison between theoretical evolutionary isochrones Z=0.002, Y=0.26, log(age)=10.15 (B08) (black)
      and the {\textsc{Teramo}} isochrones Z=0.004, Y=0.251, age=14 Gyr (red) and  Z=0.002, Y=0.248, age=14 (green).}}
    \label{Fig:iso}
  \end{figure}    
This result can be understood if we consider the color-magnitude diagram (CMD) of NGC~6652. 
It can be ascertained in the left panel of Fig.~\ref{Fig:iso} that the distribution of stars 
on the cluster CMD (\citealt{Sarajedini07}) is described well by the isochrone Z=0.002, Y=0.26, log(age)=10.15 (B08) 
for all evolutionary stages, 
except for the HB which is red in the case of NGC~6652. 
However, since there are hot luminous stars on the continuation of the cluster HB and of the Main sequence, and
some of these stars may contribute to the integrated spectrum, one may conclude that this isochrone can describe
the spectrum reasonably well. Fig.~\ref{Fig:synSp} in the Appendix illustrates a comparison 
between the original non-smoothed theoretical IL spectra computed using different isochrones. 
Fig.~\ref{Fig:Salpeter} displays the comparison between the observed continuum normalized spectrum of NGC6652 and the
 synthetic IL spectra computed using the same isochrone and elemental abundances but different MFs.
 It can be seen that there are no significant differences between the synthetic IL spectra computed applying the Ch05 and \cite{Salpeter55} MFs.

 Figures~\ref{Fig:BLs1_2} and \ref{Fig:BLs3_4} in the Appendix demonstrate how the changes in age and $ \rm Y$
values influence the synthetic profiles of three Balmer lines in comparison with the continuum normalized IL 
spectrum of NGC~6652. We considered the following isochrones by B08 for this demonstration:
(1) $ \rm Z=0.002$, $ \rm Y=0.23$, $ \rm log(age)=10.10$; (2) $ \rm Z=0.002$, $ \rm Y=0.23$, $ \rm log(age)=10.15$; (3) $ \rm Z=0.002$, $ \rm Y=0.26$, $ \rm log(age)=10.10$;
(4) $ \rm Z=0.002$, $ \rm Y=0.30$, $ \rm log(age)=10.10$. There is no isochrone $ \rm Z=0.002$, $ \rm Y=0.30$, $ \rm log(age)=10.15$ in the B08
model set. Comparison of these isochrones with the CMD of NGC~6652 is depicted in Figures~\ref{Fig:cmdBLs1_2} and \ref{Fig:cmdBLs3_4}
It can be seen in Fig.~\ref{Fig:cmdBLs1_2} that 
isochrones (1) and (2) describe all the stellar evolutionary stages reasonably well, except for the aforementioned 
blue stars on the continuation of the HB. The synthetic Balmer lines appear to be narrow and shallow in comparison with the 
observed ones (see Fig.~\ref{Fig:BLs1_2}). The same situation occurs in case we choose the isochrone (3) 
(see Figures~\ref{Fig:BLs3_4} (bottom) and \ref{Fig:cmdBLs3_4} (left)). If we choose  
isochrone (4), the synthetic Balmer lines demonstrate wider wings and deeper cores than the observed ones 
(see Figures~\ref{Fig:BLs3_4} (top) and \ref{Fig:cmdBLs3_4} (right)).

In the following we will address the question of why the {\textsc{Teramo}} isochrones can not describe the spectrum as well.
Fig.~\ref{Fig:iso} (right panel) illustrates the comparison between three isochrones in the 
"$\rm T_{eff}$ -- luminosity" plane. 
It can be ascertained that slopes of the theoretical red giant branches are slightly different in case
we choose for comparison the isochrones $ \rm Z=0.002$, $ \rm Y=0.26$, $ \rm log(age)=10.15$ (B08) and $ \rm Z=0.002$, $ \rm Y=0.248$, $\rm age=14$~Gyr 
({\textsc{Teramo}}). The ranges of $ \rm T_{eff}$ and $ \rm lg L$ for HBs are also different.
At the same time, $ \rm T_{eff}$ and $ \rm LgL$ data of other evolutionary stages agree well.
If we choose for comparison the isochrones $ \rm Z=0.002$, $ \rm Y=0.26$, $ \rm log(age)=10.15$ (B08) and $ \rm Z=0.004$, $ \rm Y=0.251$, $\rm age=14$~Gyr
({\textsc{Teramo}}), the shapes and slopes of all
evolutionary sequences, except HB, look parallel. However, the second isochrone is noticeably shifted to the lower $\rm T_{eff}$.
Additionally, the HB is cooler and less luminous in comparison with the isochrone by B08 .

\subsection{Abundances of chemical elements}
Most spectroscopic lines are blended at the resolution of $\lambda / \delta(\lambda)\sim1600$.
To determine chemical abundances, we selected dominant features in the spectra mostly sensitive to the abundances of
the specific spectroscopic elements. 
The random errors of chemical abundances determined using our method and high $\rm S/N$ observed spectra
($ \rm C/N \ge 100$) are small ( $ \rm \sigma \sim 0.1 \div 0.15$~dex)
for chemical elements with intensive absorption lines dominating the spectrum.
These elements are Fe, C, Mg and Ca. Comparison of our measurements for 20 Galactic GCs with the literature 
data obtained in high-resolution and IL spectroscopic studies indicates that the standard deviations
of the differences between our and literature estimates of $ \rm [Fe/H]$, $ \rm [Ca/Fe]$ and $ \rm [Mg/Fe]$ are $\sim$0.15~dex 
(\citealt{SS19}). 
The systematic shifts between our and literature data are $ \rm -0.06 \div -0.07$~dex for $ \rm [Ca/Fe]$ and $ \rm [Mg/Fe]$.
The systematic shift between our and literature $ \rm [Fe/H]$ data is $\sim-0.2$~dex. 
We suppose that the reason is high microturbulence velocities $ \rm \xi_{turb}$ used by us
(please, see Sec.~3.1 in \cite{SSK17} for a detailed description of the problem). 
The systematic difference of $0.37 \pm 0.17$~dex exists between our $ \rm [C/Fe]$  and literature $ \rm [C/Fe]$ values determined
considering high resolution spectra (\citealt{SS19} and references therein). We interpret this result 
as the effect of a change in the chemical composition of the atmospheres of stars in the process of their evolution
(\citealt{Kraft94}), because this difference vanishes while comparing our $ \rm [C/Fe]$ estimates with the corresponding 
literature data from IL spectroscopy (\citealt{SS19} and references therein).
The random errors of chemical abundances determined applying our method are larger for the elements with weak and
blended lines ($ \rm \sigma \sim0.15\div0.25$~dex): $ \rm Ti$, $ \rm Cr$, $ \rm Mn$ and $ \rm Na$. The typical errors for the elements 
$ \rm N$ and $ \rm O$ are $ \rm \sigma \ge 0.3$~dex. These elements have spectroscopic features too weak to be detected
in the studied spectral range. However, they influence the molecular and ionization equilibrium of other elements.
Part of the $ \rm C$ atoms forms the $ \rm CO$ molecule if the $ \rm O$ abundance is enhanced. 
This reduces the intensity of the molecular bands $ \rm CN$ and $ \rm CH$. Therefore, testing the equilibrium between 
different observed and theoretical intensities of these bands allows one to estimate abundances of all $ \rm CNO$ group
elements (\citealt{SSD13}, \citealt{SSK17}).

 Table~\ref{Tab:abund} shows the result of elemental abundances determination applying pixel-by-pixel fitting
of the observed spectrum by the synthetic one.
The determined metallicity is $\rm [Fe/H]=-1.2$~dex with $ \rm \sigma \pm0.1$~dex (random error corresponding to the internal accuracy of our method). 
Fig.~\ref{Fig:sp} features the continuum normalized IL spectrum of NGC~6652 (\citealt{Schiavon05}) 
in a wide spectral range in comparison with the synthetic IL spectrum computed using the determined abundances (Table~\ref{Tab:abund}), 
the Ch05 MF and the isochrone
Z=0.002, Y=0.26, log(age)=10.15 (B08), and smoothed to the resolution of the observed spectrum. Fig.~\ref{Fig:Conroy}
depicts the continuum normalized spectrum of NGC~6652 (\citealt{Schiavon05}) (black) 
in comparison with the synthetic IL spectrum (magenta) computed utilizing  the isochrone Z=0.002, Y=0.26, log(age)=10.15 
   (B08), $ \rm [Fe/H]=-0.93$~dex, elemental abundances listed in Table\ref{Tab:abund} and the MF by Ch05.
\begin{table}
\begin{center}
\caption[]{Estimated abundances $\rm [X/Fe]$ in dex and abundance errors. 
Abundances from \citealt{Conroy} (C18) are provided for comparison.}\label{Tab:abund}
\scriptsize{
 \begin{tabular}{lccccccccc}
  \hline\noalign{\smallskip}
Ref./Elem. &      C             & N            &  O           & Na            & Mg        & Ca              & Ti           & Cr           & Mn          \\
  \hline\noalign{\smallskip}
ours       &      0.20$\pm$0.14 & 0.30$\pm$0.25& 0.3$\pm$0.25 & 0.45$\pm$0.20 & 0.40$\pm$0.10 & 0.30$\pm$0.10& -0.10$\pm$0.15& -0.35$\pm$0.2 & -0.35$\pm$0.23 \\
C18        & 0.02$\pm$0.02      & 0.23$\pm$0.05& --           &   --          & 0.45$\pm$0.02 & 0.25$\pm$0.03& 0.25$\pm$0.03&  --          &      --      \\
  \noalign{\smallskip}\hline
\end{tabular}}
\end{center}
\end{table} 
\section{Discussion}
 As was mentioned in the introduction, the origin of NGC~6652 is a matter of debate. 
According to the distance to the Galactic center, metallicity and age, this GC belongs to the bulge and has formed in-situ 
(\citealt{Bica16}, \citealt{Massari19}). 
The origin of the bulge itself is not completely understood (\citealt{Barbuy18} and references therein). 
Stellar populations with chemical and kinematic properties from different Galactic subsystems are mixed there.
The metallicity of NGC~6652 corresponds well to the peak of $ \rm [Fe/H]$ values for the majority of GCs in the bulge: $ \rm -1.3\le[Fe/H]\le-0.8$, 
a metallicity peak similar to that of RR Lyrae stars. Without any doubt, this GC is one of the oldest in our Galaxy. 

Prototypical bulge stars and GCs are enhanced in $\alpha$ \rm -process elements: $ \rm O$, $ \rm Mg$, $ \rm Ci$, and $ \rm Ca$ (\citealt{Barbuy18}).
The abundances of $ \rm Mg$ and $ \rm Ca$ confidently determined by us and by \cite{Conroy} correspond well to the values typical 
for bulge stellar populations with $ \rm [Fe/H]\sim-1$~dex.

Sodium is among odd-Z elements produced together with $ \rm Mg$ in massive stars. Its abundance is high in bulge stars (\citealt{Barbuy18}).
Our estimated abundance of $ \rm Na$ is also high.

Manganese and chromium are Fe-peak elements. Low $ \rm [Mn/Fe]$ and $ \rm [Cr/Fe]$ values obtained by us are in line with the measurement 
for bulge stars (\citealt{Barbuy18}). 

Titanium is an iron peak element, but its enhancement depending on metallicity looks similar to that of $\alpha$ \rm -process elements 
(e.g. \citealt{Barbuy18}).
Our $ \rm [Ti/Fe]$ value is too low in comparison to the value determined by \cite{Conroy}. This element does not have
intensive lines in the studied spectral range. Probably, our data is not accurate. 
However, let us note that there are other Galactic GCs in the bulge containing stars with reasonably low $ \rm [Ti/Fe]$:
NGC~6522, NGC~6528, NGC~6553, NGC~6558 (Table~3 in \citealt{Bica16}). Some of this objects reveal moderate $\alpha$ \rm -element
enhancement. Galactic field RR Lyrae stars with $ \rm [Fe/H]\ge-1.0$~dex have unusually low $ \rm [Ti/Fe]$ 
(\citealt{Marsakov19} and references therein). Low $ \rm [Ti/Fe]$ values are characteristic of stellar populations in dwarf galaxies 
(e.g. \citealt{Carlin18} and references therein). Future studies of $ \rm Ti$ stellar abundances in NGC~6652 will definitely 
help to improve the data.

\section{Conclusions}
\label{sect:conclusion}
We use IL spectrum of the Galactic GC NGC~6652 (\citealt{Schiavon05}) and our method (see e.g. \citealt{SS19}, \citealt{SSK18}, \citealt{SSK17}) 
to derive the age $\sim13.6\pm1$~Gyr, $\rm Y=0.26\pm0.2$, $\rm[Fe/H]=-1.2\pm0.1$~dex and
 abundances of chemical elements: $ \rm C$, $ \rm N$, $ \rm O$, $ \rm Na$, $ \rm Mg$, $ \rm Ca$, $ \rm Ti$, $ \rm C$ and $ \rm Mn$.
We conclude that the Ch05 MF and the isochrone Z=0.002, Y=0.26, log(age)=10.15 (B08) 
allow us to reproduce well the shapes and intensities of the Balmer absorption lines in the observed spectrum.
The determined metallicity is $0.2-0.3$~dex lower than the literature value (\citealt{ZinnWest84}).
This systematic deviation is specific for our analysis (\citealt{SS19}). We argue that several hot HB stars   
contribute to the spectrum of NGC~6652 (\citealt{Schiavon05}). 
 The derived elemental abundances agree with the corresponding abundances of bulge stars and GCs (\citealt{Bica16}, \citealt{Barbuy18}). 
The only exclusion is $ \rm [Ti/Fe]$. The lines of this element are weak and blended at the examined spectral range and resolution.
Future high-resolution spectroscopic studies will definitely help 
to establish the chemical composition of stars in NGC~6652 and the origin of this GC. 

\begin{acknowledgements}
We thank an anonymous referee for the report which helped to improve the paper. 
This work is supported by the RFBR grant No. 18-02-00167 a.
\end{acknowledgements}

\newpage
\appendix                 
\section{Integrated-light synthetic spectra computed using different isochrones and different mass functions}
In this section, we illustrate the comparison between synthetic spectra computed using 
our method (see \citealt{SS19}, \citealt{SSK18}, \citealt{SSK17} and references therein), elemental abundances listed in Table\ref{Tab:abund}, 
and different isochrones and
the comparison between IL spectrum of NGC~6652 (\citealt{Schiavon05}), and the synthetic spectra computed 
applying different MFs: the Ch05 (equation~2 in that paper) 
and the \cite{Salpeter55} MFs.

     \begin{figure}[h]
   \centering
   \includegraphics[bb=53 30 570 800,width=7cm, angle=-90,clip]{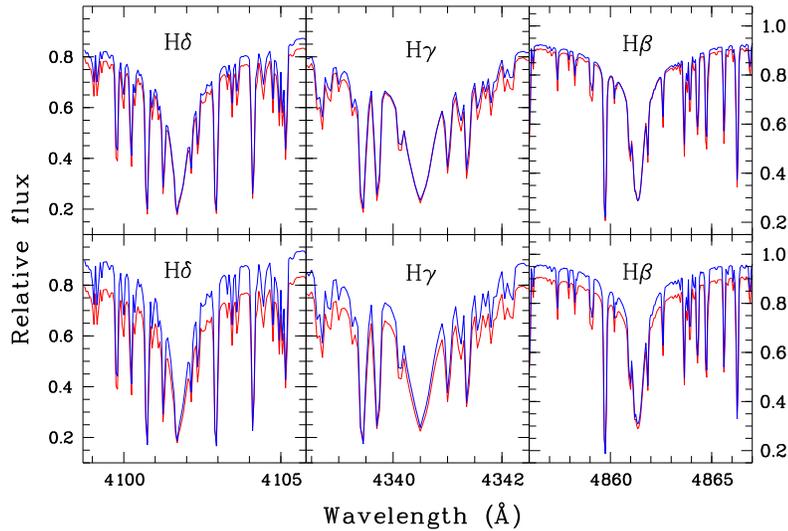}
   \caption{{\small Comparison of three Balmer line profiles in the synthetic high-resolution spectra computed using the MF by 
   Ch05, the same elemental abundances and different isochrones: 
   the isochrone by B08 Z=0.002, Y=0.26, log(age)=10.15 displayed
   in red in the top and bottom panels and the isochrones by the {\textsc{Teramo}} group 
   (\citealt{Pietrinferni+13} and references therein): Z=0.004, Y=0.251, age=14 Gyr and 
    Z=0.002, Y=0.248, age=14 Gyr depicted in blue in the bottom and top panels, respectively.}} 
   \label{Fig:synSp}
   \end{figure}
Fig.~\ref{Fig:synSp} illustrates the comparison 
between the original non-smoothed theoretical IL spectra computed using different isochrones:
the isochrone by B08 Z=0.002, Y=0.26, log(age)=10.15 signified
   in red in the top and bottom panels, the isochrones by the {\textsc{Teramo}} group 
   (\citealt{Pietrinferni+13} and references therein) Z=0.004, Y=0.251, age=14 Gyr and 
    Z=0.002, Y=0.248, age=14 Gyr depicted in blue in the bottom and top panels, respectively.
     
    \begin{figure}[h]
   \centering
   \includegraphics[bb=53 30 570 800,width=7cm, angle=-90,clip]{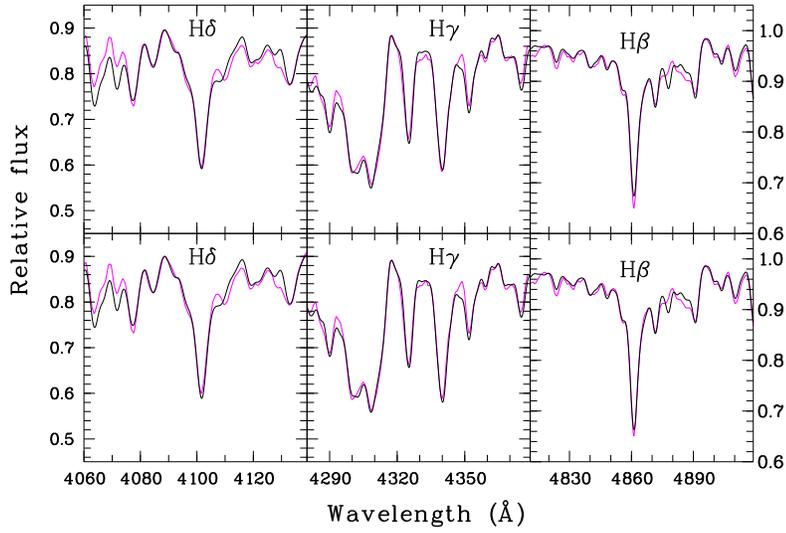}
   \caption{{\small Comparison of three Balmer lines in the continuum normalized spectrum of NGC~6652 (\citealt{Schiavon05}) (magenta) 
   and the synthetic IL spectra (black) computed using the same isochrone Z=0.002, Y=0.26, log(age)=10.15 
   (B08) and elemental abundances but different MFs: the MF by Ch05 (bottom)
   and the MF by \cite{Salpeter55} (top).}}
   \label{Fig:Salpeter}
   \end{figure}
 Fig.~\ref{Fig:Salpeter} shows a comparison between the observed continuum normalized spectrum of NGC6652 and the
 synthetic IL spectra computed using the same isochrone and elemental abundances but different MFs.
 The synthetic spectra were smoothed to the resolution of the observed spectrum. 
 It can be seen that the comparison is good in both cases and that 
 there are no significant differences between the two synthetic IL spectra.
    
     \begin{figure}[h]
   \centering
   \includegraphics[width=9cm, angle=-90]{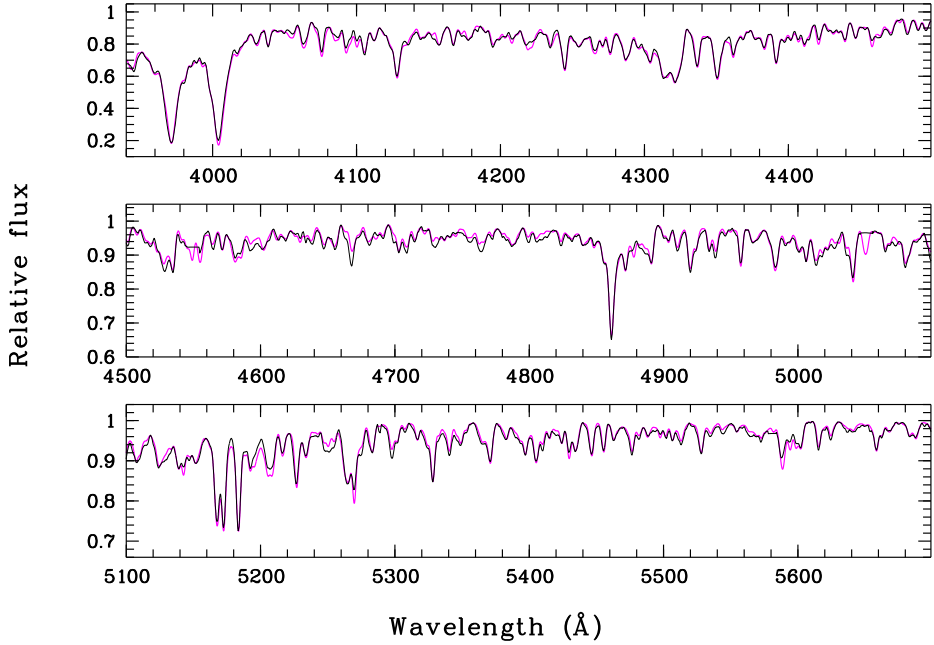}
   \caption{{\small Comparison of the continuum normalized spectrum of NGC~6652 (\citealt{Schiavon05}) (black) 
   and the synthetic IL spectrum (magenta) computed using  the isochrone Z=0.002, Y=0.26, log(age)=10.15 
   (B08), the elemental abundances listed in Table\ref{Tab:abund} and the MF by Ch05.}}
   \label{Fig:sp}
   \end{figure}   
  \begin{figure}[h]
   \centering
   \includegraphics[width=9cm, angle=-90]{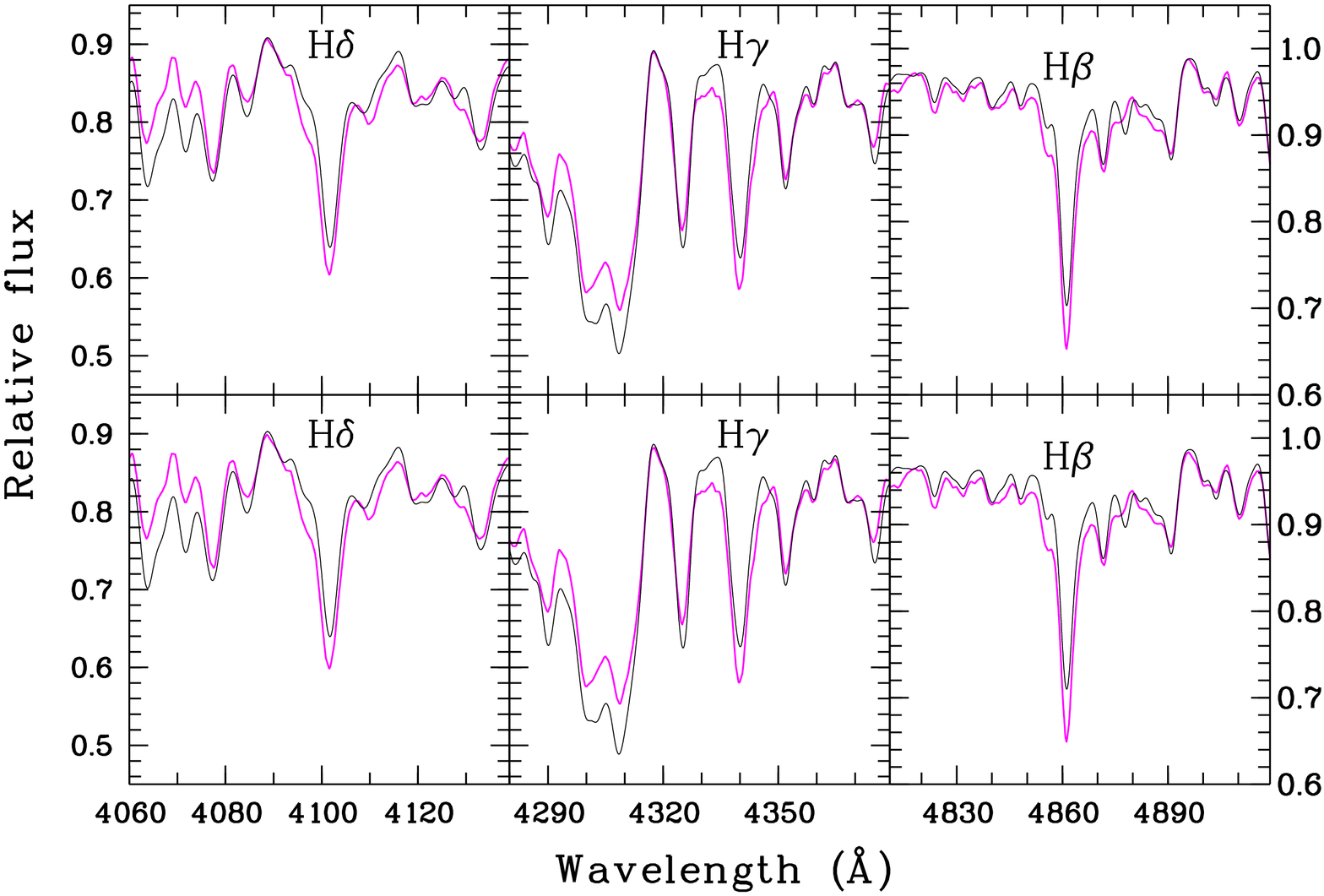}
   \caption{{\small Comparison of three Balmer lines in the continuum normalized spectrum of NGC~6652 (Schiavon et al. 2005)
(black) and the synthetic IL spectra (magenta) computed using the MF by Ch05, the same elemental
abundances and different isochrones (B08): Z=0.002, Y=0.23, log(age)=10.15 (bottom) and
Z=0.002, Y=0.23, log(age)=10.10 (top).}}
   \label{Fig:BLs1_2}
   \end{figure}    
     \begin{figure}[h]
   \centering
   \includegraphics[width=9cm, angle=-90]{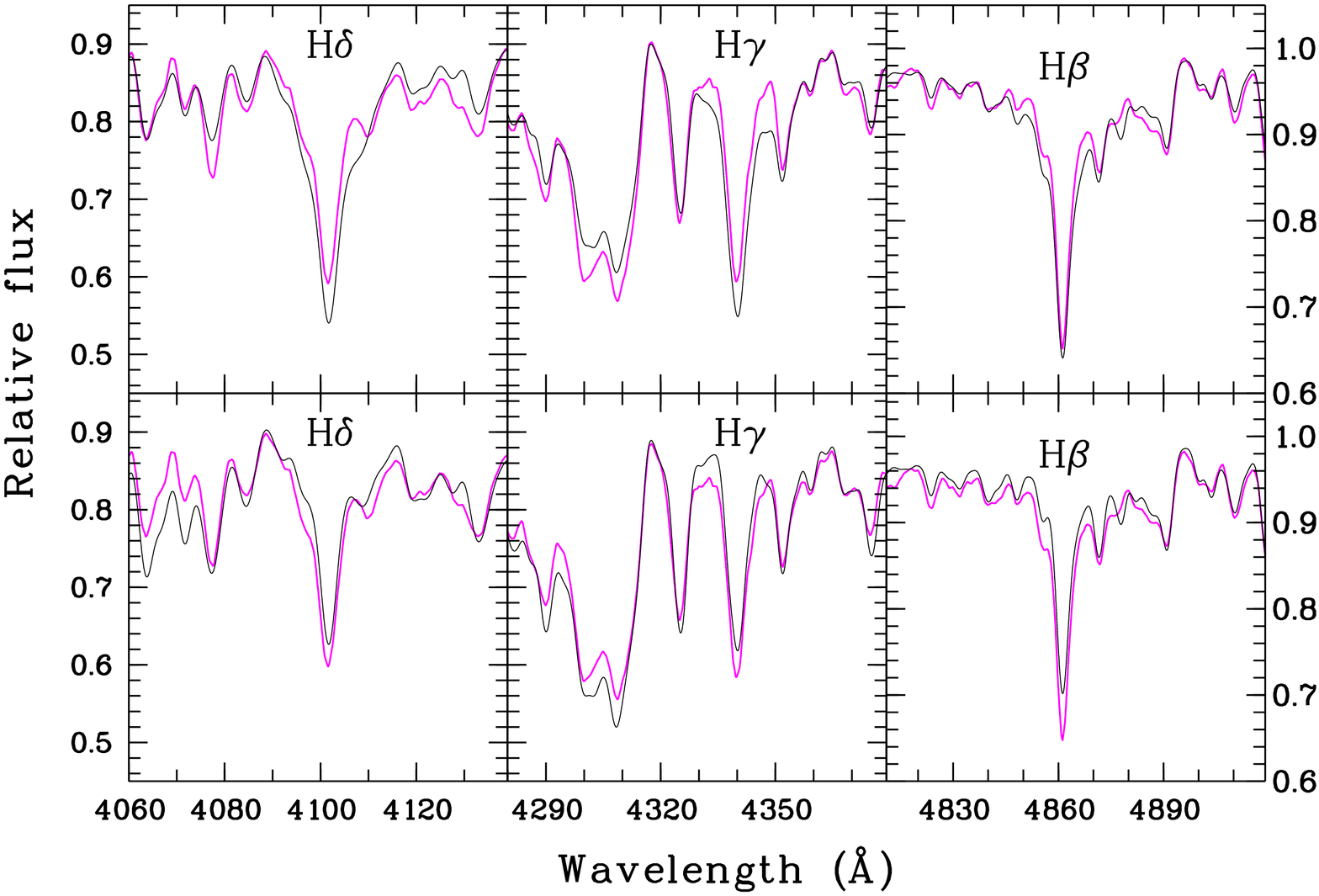}
   \caption{{\small Comparison of three Balmer lines in the continuum normalized spectrum of NGC~6652 (Schiavon et al. 2005)
(black) and the synthetic IL spectra (magenta) computed using the MF by Ch05, the same elemental
abundances but different isochrones (B08): Z=0.002, Y=0.26, log(age)=10.10 (bottom) and
Z=0.002, Y=0.30, log(age)=10.10 (top).}}
   \label{Fig:BLs3_4}
   \end{figure}   
     \begin{figure}[h]
   \centering
   \includegraphics[width=9cm, angle=-90]{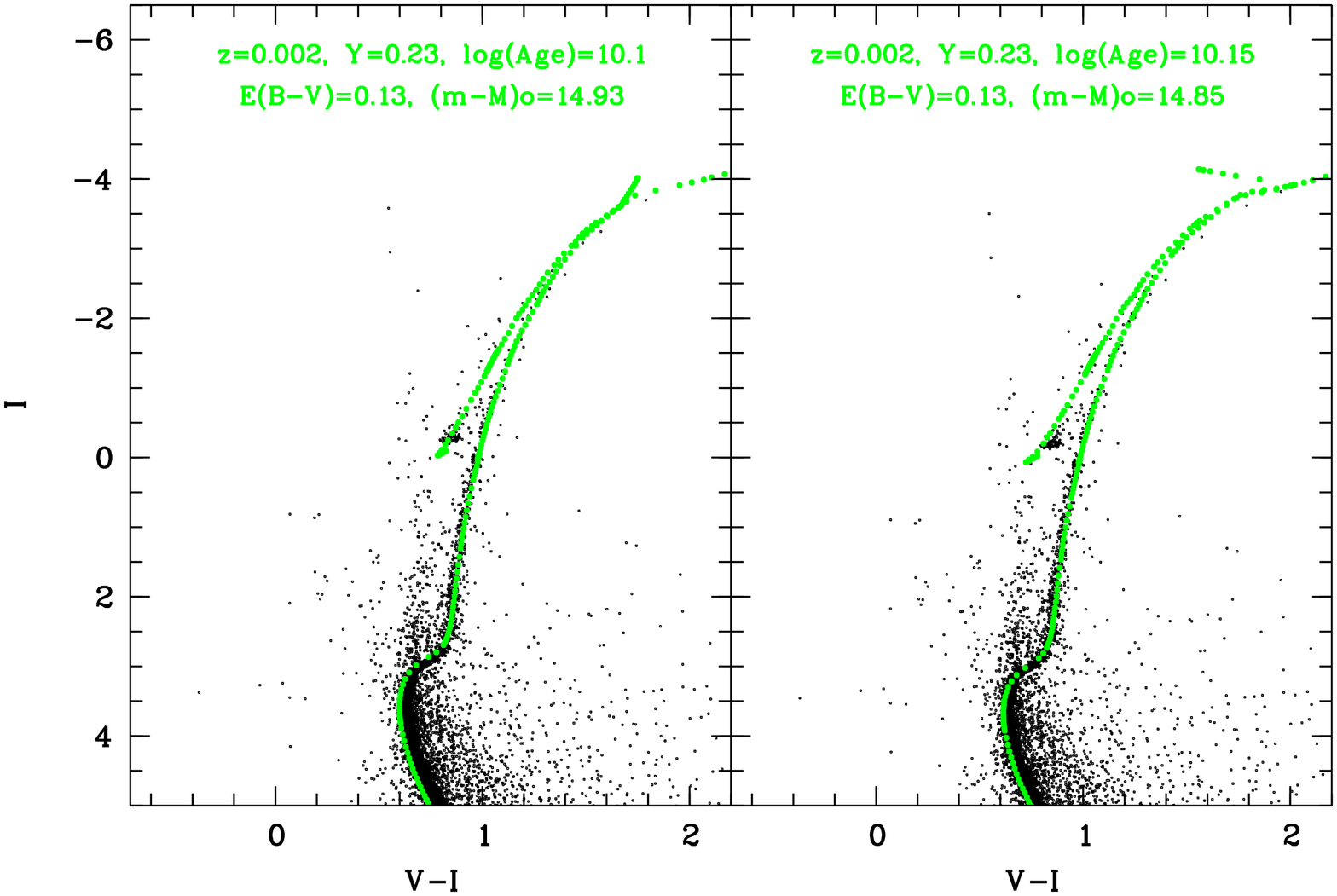}
   \caption{{\small Comparison between the CMD of NGC6652 (Sarajedini et al. 2007) and the theoretical evolutionary
isochrones (B08): Z=0.002, Y=0.23, log(age)=10.10 (left) and
Z=0.002, Y=0.23, log(age)=10.15 (right).}}
   \label{Fig:cmdBLs1_2}
   \end{figure} 
 \begin{figure}[h]
   \centering
   \includegraphics[width=9cm, angle=-90]{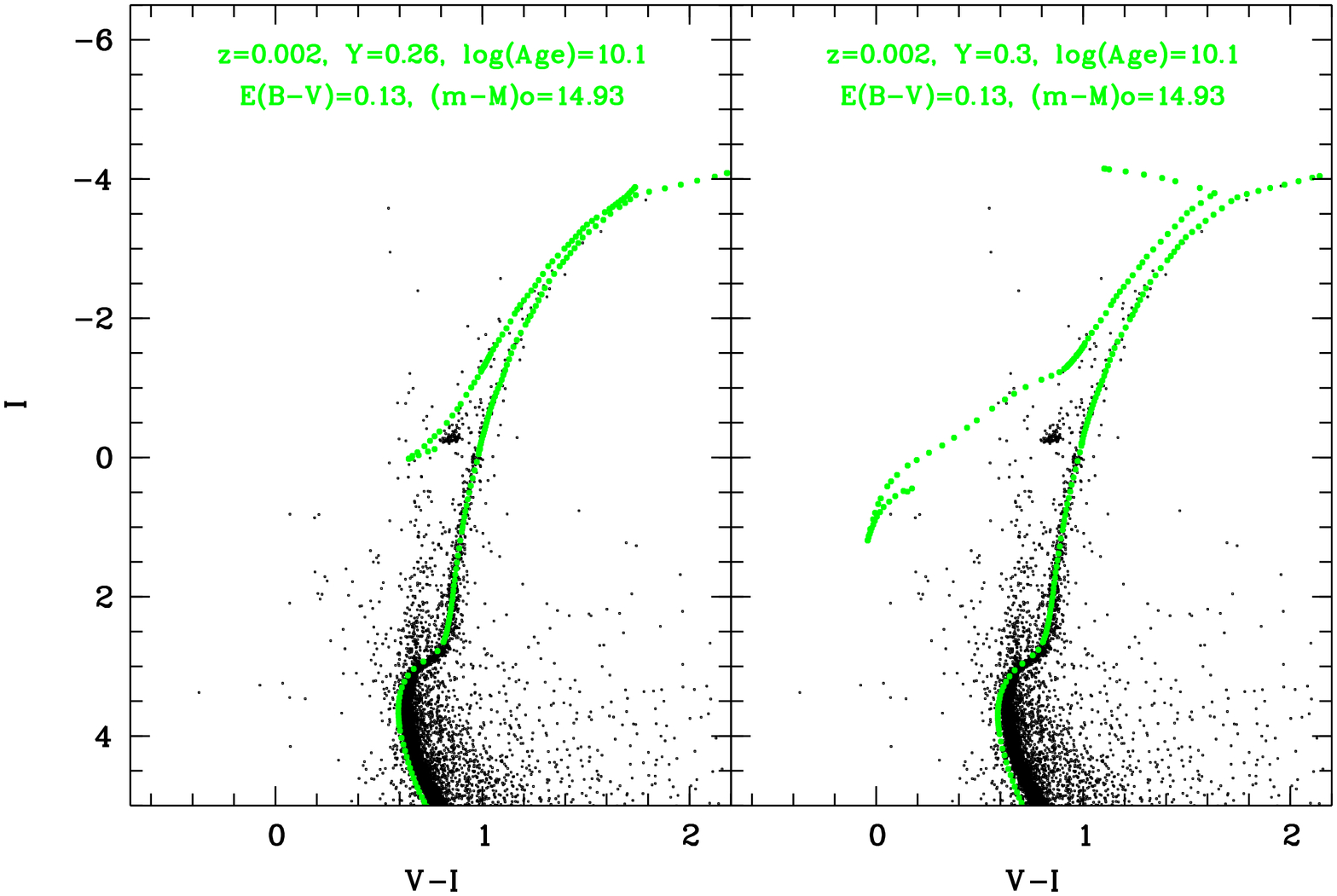}
   \caption{{\small Comparison between the CMD of NGC6652 (Sarajedini et al. 2007) and the theoretical evolutionary
isochrones (B08): Z=0.002, Y=0.26, log(age)=10.10 (left) and
Z=0.002, Y=0.30, log(age)=10.10 (right).}}
   \label{Fig:cmdBLs3_4}
      \end{figure}
      
   \begin{figure}[h]
   \centering
   \includegraphics[width=9cm, angle=-90]{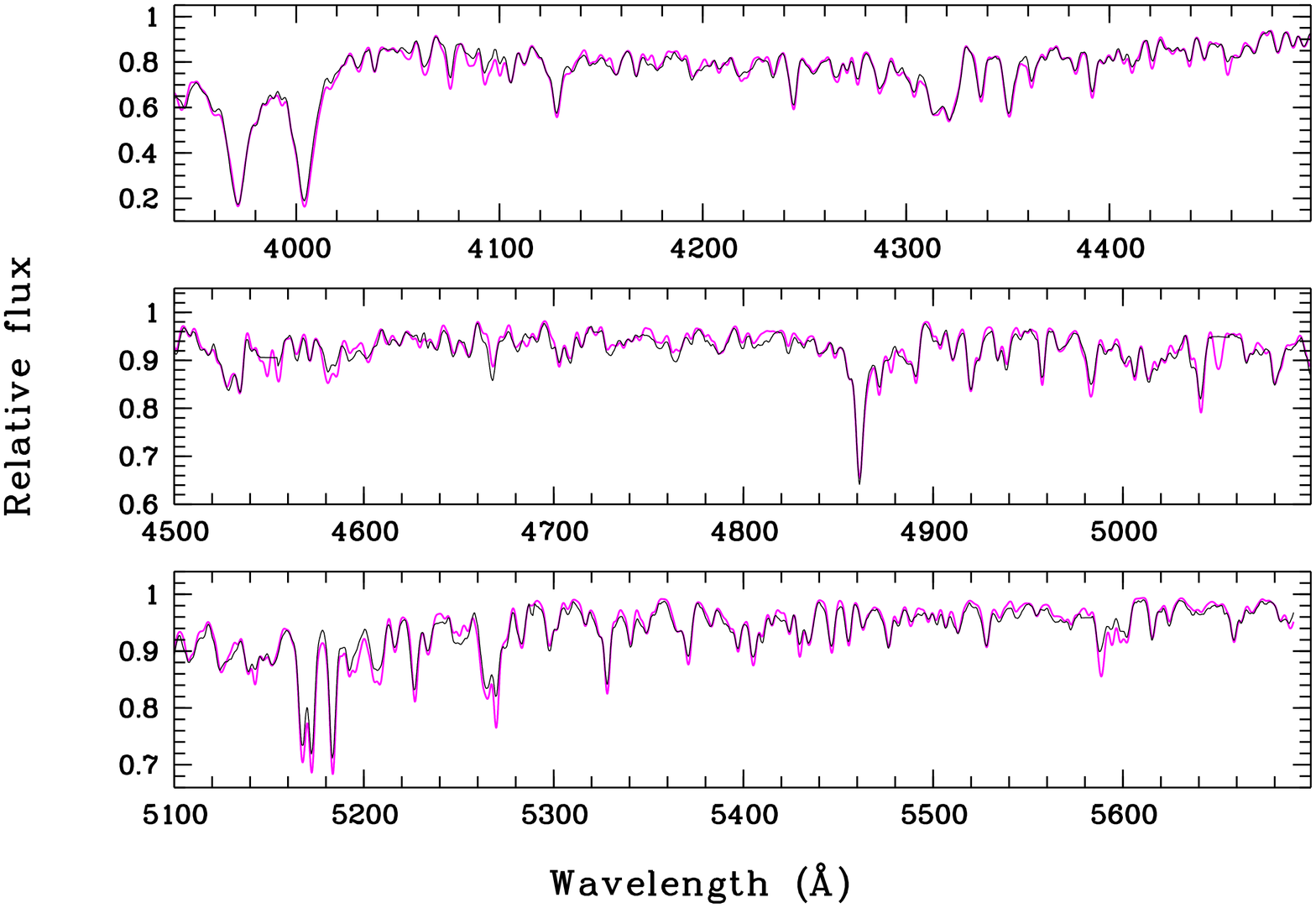}
   \caption{{\small Comparison of the continuum normalized spectrum of NGC~6652 (\citealt{Schiavon05}) (black) 
   and the synthetic IL spectrum (magenta) computed using  the isochrone Z=0.002, Y=0.26, log(age)=10.15 
   (B08), $ \rm [Fe/H]=-0.93$~dex, elemental abundances listed in Table\ref{Tab:abund} and the MF by Ch05.}}
   \label{Fig:Conroy}
   \end{figure}
   
   Fig.~\ref{Fig:sp} features the continuum normalized IL spectrum of NGC~6652 (\citealt{Schiavon05}) 
 in comparison with the synthetic IL spectrum computed utilizing the elemental abundances listed in Table\ref{Tab:abund},
 the Ch05 MF and the isochrone
Z=0.002, Y=0.26, log(age)=10.15 (B08) and smoothed to the resolution of the observed spectrum.
 To approximate the shape of the continuum in the observed spectra, we first smoothed the spectra by replacing 
 each pixel value by the maximum of all points in the window $ 2\times \rm FWHM*10 + 1$, where $\rm FWHM$ is the average full width at 
 half maximum for spectroscopic lines in the observed spectrum. Then, we 
 implemented the running mean with the radius of $10\times \rm FWHM$ to the filtered spectrum.
 
 Figures~\ref{Fig:BLs1_2} and \ref{Fig:BLs3_4} show the synthetic IL spectra computed in similar way as was
 described in the previous paragraph, but using the isochrones: Z=0.002, Y=0.23, log(age)=10.15 (Fig.~\ref{Fig:BLs1_2}, bottom);
Z=0.002, Y=0.23, log(age)=10.10 (Fig.~\ref{Fig:BLs1_2}, top); Z=0.002, Y=0.26, log(age)=10.10 (Fig.~\ref{Fig:BLs3_4} bottom) and
Z=0.002, Y=0.30, log(age)=10.10 (Fig.~\ref{Fig:BLs3_4}, top).
The comparison between the CMD of NGC6652 (Sarajedini et al. 2007) and the aforementioned four theoretical evolutionary
isochrones (B08) is depicted in Figures~\ref{Fig:cmdBLs1_2} and \ref{Fig:cmdBLs3_4}.

 Fig.~\ref{Fig:Conroy} displays what happens if we apply $ \rm [Fe/H]=-0.93$~dex for the synthetic spectral calculation, the value obtained by \cite{Conroy}.
 Other elemental abundances are listed in Table\ref{Tab:abund}.

   \label{lastpage}
\end{document}